\documentclass[reprint, prl, superscriptaddress, nofootinbib, preprintnumbers]{revtex4-1}

\pdfoutput=1
\usepackage{mathtools, amsfonts, amsthm, latexsym, amssymb}

\usepackage{color}
\usepackage{hyperref}
\hypersetup{linktocpage, colorlinks=true, linkcolor=blue, citecolor=blue, urlcolor=blue}

\usepackage{slashed}

\DeclareMathOperator{\re}{Re}
\renewcommand{\L}{\mathcal L}

\newcommand{\alignStart}{ \begin{equation} \begin{aligned} }
\newcommand{\alignEnd}{ \end{aligned} \end{equation} }
\newcommand{\gatherStart}{ \begin{equation} \begin{gathered} }
\newcommand{\gatherEnd}{ \end{gathered} \end{equation} }

\newcommand{\vev}[1]{{\langle #1 \rangle}}
\newcommand{\gev}{\, \text{GeV}}

\newcommand{\mpl}{M_\text{pl}}
\newcommand{\feff}{f_\text{eff}}
\begin{document}
\title{Natural Inflation and Quantum Gravity}
\author{Anton de la Fuente}
\affiliation{Department of Physics, University of Maryland, College Park, Maryland 20742, USA}%
\author{Prashant Saraswat}
\affiliation{Department of Physics, University of Maryland, College Park, Maryland 20742, USA}%
\affiliation{Department of Physics and Astronomy, Johns Hopkins University, Baltimore, Maryland 21218, USA}%
\author{Raman Sundrum}
\affiliation{Department of Physics, University of Maryland, College Park, Maryland 20742, USA}%
\date{\today}

\preprint{UMD-PP-014-023}
\begin{abstract}
Cosmic Inflation provides an attractive framework for understanding the early universe and the cosmic microwave background. It can readily involve energies close to the scale at which Quantum Gravity effects become important. General considerations of black hole quantum mechanics suggest nontrivial constraints on any effective field theory model of inflation that emerges as a low-energy limit of quantum gravity, in particular the constraint of the Weak Gravity Conjecture. We show that  higher-dimensional gauge and gravitational dynamics can elegantly satisfy these constraints and lead to a viable, theoretically-controlled and predictive class of Natural Inflation models. 
\end{abstract}


\maketitle
The success of modern cosmology is founded on the simplifying features of homogeneity, isotropy and spatial flatness of the Universe on the largest distances. In this limit, spacetime evolution is given in terms of a single scale-factor, $a(t)$, and its Hubble expansion rate, 
$H \equiv \dot{a}/a$. Homogeneity and flatness are themselves puzzling, constituting 
 very special ``initial'' conditions from the viewpoint of the Hot Big Bang (HBB). 
 But they become more robust if the HBB is preceded by an even earlier era of Cosmic Inflation, exponential expansion of the Universe driven by the dynamics of a scalar field $\phi$ (the ``inflaton'') coupled to General Relativity (see \cite{baumann} for a review): 
\begin{gather}
	H^2 = \frac{8\pi G_N}{3} \left[\tfrac{1}{2} \dot \phi^2 + V(\phi) \right]  \nonumber \\
	\ddot \phi + 3 H \dot \phi + V' = 0.
\end{gather}
(We work in fundamental units in which $\hbar = c =1$. $G_N$ is Newton's constant.) 
If ``slow roll'' is achieved for a period of time, $\dot{\phi}$ subdominant and $V(\phi) \approx$ constant, we get $a \propto e^{H t}, H \approx$ constant, after which the potential energy is released, ``reheating'' the Universe to the HBB. 
Phenomenologically, ${\cal N}_\text{e-folds}  > 40 - 60$  are required to understand the degree of homogeneity/flatness we see today. 

Remarkably, quantum fluctuations during inflation can seed the inhomogeneities in the distribution of galaxies and in the Cosmic Microwave Background (CMB). In particular, the CMB temperature-fluctuation  power-spectrum, 
\begin{equation}
\Delta_S(k) \propto  k^{n_s(k)}, k \equiv {\rm wavenumber}, n_s \equiv {\rm spectral~index}, 
\end{equation}
is generically predicted by inflation to be approximately scale-invariant, $n_s \approx 1$, and is measured to be $n_s \approx 0.96$ \cite{wmap, planck}.

Slow roll itself requires an unusually flat potential, suggesting that the inflaton $\phi$ is a pseudo-Nambu-Goldstone boson of a spontaneously broken global $U(1)$ symmetry, an ``axion". If there is a weak coupling that 
 \emph{explicitly} violates $U(1)$ symmetry by a definite amount of charge,  one can generate a potential,
\begin{equation}
	V(\phi) = V_0 \left(1-\cos \frac{\phi}{f} \right),
\label{eq:natural}
\end{equation}
where $f$ is a constant determined by the spontaneous breaking dynamics, while $V_0$ is a constant proportional to the weak coupling.
This is the model of ``Natural Inflation''~\cite{FreeseFriemanOlinto}.\footnote{The fine-tuning of the two terms in Eq.~\eqref{eq:natural} to obtain a 
(nearly) vanishing vacuum energy relates to the notorious Cosmological Constant Problem~\cite{ccreview}, which we do not address here.} It can be successfully fit to data, and in particular for ${\cal N}_\text{e-folds} >  50, n_s \approx 0.96$, one finds~\cite{planck}
\gatherStart
           f  >  2 \times 10^{19} \gev \approx 10  \mpl \\
	V_0  > (2 \times 10^{16} \gev)^4 \approx (10^{-2} \mpl)^4.	
\label{eq:data}
\gatherEnd
The Planck scale $\mpl \equiv 1/\sqrt{8\pi G_N} = 2 \times 10^{18} \gev$ is the energy scale
above which Quantum Gravity (QG) effects become strong, and   
effective field theory (EFT) must break down in favor of a more fundamental description such as superstring theory \cite{Polchinski}.  


The very high energy scale $V_0^{1/4} \approx 0.01 \mpl$ is without precedent in observational physics and implies sensitivity to new exotic phenomena. For such large inflationary energy densities, quantum graviton production during inflation gives rise to a tensor/scalar ratio of the CMB power spectrum of $r \sim 0.1$. Indeed, the BICEP2 CMB experiment has claimed a signal at this sensitivity \cite{bicep2}, although there is still serious concern over possible contamination by foreground dust~\cite{FlaugerDust, PlanckDust}.  Regardless, cosmological observations have the potential to provide information about physics at the highest energy scales in the near future.

However, the proximity of the QG scale raises concerns about the validity of effective field theory treatments of inflation and susceptibility to poorly-understood QG effects. There are broadly two approaches to addressing such QG uncertainties in high-scale inflation models. One is to derive inflationary models directly within known superstring constructions, which provide reasonably explicit UV descriptions of QG. 
Such constructions feature many moduli fields (for example, describing the size and shape of several extra dimensions) which must be stabilized and which also receive time-dependent back-reaction effects during the course of inflation. Consistently constructing and analyzing models of this type can be an involved and difficult task, and there is as yet no fully realistic top-down derivation. Nevertheless, considerable qualitative progress has been made on possible shapes and field-ranges of inflaton potentials in string theory and their effects~\cite{eva1,eva2,Blumenhagen:2014gta,Grimm:2014vva,Long:2014dta,eva3,Gao:2014uha,Ben-Dayan:2014lca,Kenton:2014gma}.

Alternatively, one can try to construct bottom-up effective field theory models, incorporating simple mechanisms that shield the inflationary structure from unknown QG corrections, aspects of which have been previously explored in e.g.~\cite{ extraNatural, Dimopoulos:2005ac, Kaloper:2008fb, Kaloper:2011jz, Dubovsky:2011tu,Harigaya:2012pg, Harigaya:2014eta,  Dine:2014hwa,Yonekura:2014oja,  Neupane:2014vwa}. However, studies of robust quantum properties of large black hole solutions in General Relativity, 
as well many string theory precedents, 
strongly suggest that there are non-trivial constraints on effective field theory couplings in order for them to be consistent with {\it any} UV completion incorporating QG, which make inflationary model-building challenging. In this paper, we will discuss the impact of such black-hole/QG considerations in the context of Natural Inflation, in particular the role of the Weak Gravity Conjecture (WGC)~\cite{WGC}. 
 While these considerations rule out some inflationary models,  we demonstrate for the first time that there do exist simple and predictive effective theories of natural inflation, consistent with the WGC, where the inflaton arises from components of higher-dimensional gauge fields.  The advantage of the effective field theory approach is two-fold: (i) the models have relatively few moving parts, whose dynamics can be analyzed quite straightforwardly and comprehensively, and (ii) one can achieve full realism. We believe that such a higher-dimensional realization yields the most attractive framework for cosmic inflation to date. Further elaboration of our work will be presented in \cite{followUp}. 


\section{Quantum Gravity Constraints}
Classical black holes can carry gauge charges, observable by their gauge flux outside the horizon, but not global charges. 
Studies of black hole formation and Hawking evaporation, combined with the statistical interpretation of their entropy, 
then imply that  such quantum processes 
violate global charge conservation~\cite{susskindGlobal, Banks:2010zn}. By the Uncertainty Principle this holds even for virtual black holes, implying that at some level global symmetries such as those desired for Natural Inflation cannot co-exist with QG. Of course, global symmetries are seen in a variety of experimental phenomena, but these are accidental or emergent at low energies, while Natural Inflation only achieves slow roll for $f > \mpl$! A loop-hole is that $1/f$ may represent a weak coupling and low-scale symmetry breaking rather than very high scale breaking. The mechanism of ``Extranatural Inflation" \cite{extraNatural} precisely exploits this loop-hole,   realizing $\phi$ as a low-energy remnant of a $U(1)$ \emph{gauge} symmetry.  The model is electrodynamics, but in  4+1-dimensional spacetime, with the usual dimensions, $x^{\mu = 0-3}$, augmented by a very small extra-dimensional circle, $x^5 \in (- \pi R, \pi R]$. The 3+1-dimensional inflaton is identified with the phase of the gauge-invariant Wilson loop around the circle,
\begin{gather}
	\phi(x^{\mu}) \equiv \frac{1}{2\pi R} \oint dx^5 A_5(x^{\mu}, x^5). 
\label{eq:WilsonLoop}
\end{gather}
Classically, the masslessness of the Maxwell field, $A_{M = \mu, 5}$, matches onto $V(\phi) = 0$ in the long distance effective theory $\gg R$. 
But 4+1D charged matter, with charge $g_5$, mass $m_5$, and spin $S$, corrects the quantum effective potential \cite{hosotani,massiveHosotani}, 
\begin{gather}
	\delta V(\phi) = \frac{3(-1)^S}{4 \pi^2} \frac{1}{(2\pi R)^4} \sum_{n \in \mathbb Z}  c_n
	e^{-2\pi n R m_5}   \re e^{i n \phi/f}  \nonumber \\ 
	c_n(2\pi R m_5) = \frac{(2 \pi R m_5)^2}{3 n^3} + \frac{2 \pi R m_5}{n^4} + \frac{1}{n^5}, 
\label{eq:aharanovBohm} 
\end{gather}
where $(e^{-2\pi R m_5})/R^4$ is a typical (Yukawa-suppressed) extra-dimensional Casimir energy density, and the phase captures an
Aharonov-Bohm effect around the circle. We have written this in terms of the emergent scale,
\begin{equation}
	f \equiv \frac{1}{2\pi R g},
\label{eq:f}
\end{equation}
where $g$ is the effective 3+1 coupling which matches onto $g_5$ in the UV.
We see that Natural Inflation structure (with innocuous harmonics), with $f  > \mpl$,  can emerge at a sub-Planckian compactification scale, $1/R \ll \mpl$,  by choosing weak gauge coupling $g \ll 1$. After reaching the minimum of its potential the inflaton can ``reheat'' the Universe to a radiation-dominated phase by decaying into the charged matter.

The requirement $g \ll 1$ seems suspiciously close to $g = 0$, the limit in which the $U(1)$ gauge symmetry effectively reverts to an exact global symmetry, at odds with QG. Indeed, Extranatural Inflation runs afoul of a subtle QG criterion known as the Weak Gravity Conjecture~\cite{WGC}.  (For related work see e.g.~\cite{Li:2006vc,Li:2006jja,Banks:2006mm,Huang:2006pn,Huang:2007st,Cheung:2014vva,Cheung:2014ega}.) The WGC again uses universal features of black holes to provide insights into QG constraints on EFT. In brief, one argument is as follows. (We will discuss this and other motivations for the WGC at greater length in \cite{followUp}.) 
 Ref.~\cite{Banks:2010zn} has shown that in EFTs containing both a Maxwell gauge field and General Relativity, the associated gauge group must be {\it compact} $U(1)$, in the sense that electric charges 
 must be quantized in integer multiples of the coupling $g$, in order to avoid  other exact {\it global} symmetries and related negative consequences. 
 Then, there exist large black hole solutions to the Einstein-Maxwell Equations carrying both electric and magnetic charges. These solutions are quantum-mechanically consistent if they obey the Dirac quantization condition, whereby magnetic charges are quantized in units of $2\pi/g$. 

All precedent in General Relativity and string theory research (e.g.\@ \cite{Bekenstein, Susskind:1993ws, StromingerVafa, Wall}) suggests that black holes are themselves coarse-grained EFT descriptions of gravitational bound states of more basic components (see \cite{Peet,Sen} for reviews). In particular, magnetically charged black holes should be ``made out of" fundamental magnetic charges which are themselves not black holes. And yet, this is impossible for sufficiently small $g \ll 1$. The reason is that 
Maxwell EFT cannot described electric and magnetic charges which are both light and pointlike. Instead, the magnetic charges must be heavy solitons, with a size $1/\Lambda$, where $\Lambda < \mpl$ is the UV energy cutoff of the EFT.  The magnetostatic self-energy  in the region outside the $1/\Lambda$-sized ``core'', where EFT applies,  is then $\pi \Lambda/(2g^2) \gg \Lambda$;~\footnote{By comparison, for weakly coupled electrically-charged point particles, the length scale that sets the electrostatic self-energy is played by the Compton wavelength, which is then a small perturbation of the mass, $g^2 m /(8\pi) \ll m$.} the mass $m_\text{core}$ within the core is expected to be at least comparable to this. In order for the soliton to be larger than its horizon radius $2 G_N m_\text{core}$,  to avoid being a black hole itself, we must have
  \begin{equation}
	\Lambda \lesssim 2\sqrt{2} g \mpl.
\label{eq:WGC}
\end{equation}
Here the ``$\lesssim$" reminds us of the  ${\cal O}(1)$ uncertainties in this argument. This is the WGC. When testing theories of inflation for {\it parametric control} these ${\cal O}(1)$ uncertainties will be irrelevant,  but we will be subject to them when fitting models to precision data.

Requiring the compactification scale to be below the EFT cutoff,  $1/R < \Lambda$, then implies $f < \mpl$, by Eq.~\eqref{eq:f}, spoiling minimal Extranatural Inflation~\cite{WGC, Rudelius:2014wla}. Note that even with the ${\cal O}(1)$ uncertainty in the WGC, we cannot get parametrically large $f/\mpl$ (ie. large
 ${\cal N}_{\rm e-folds})$.



\section{Bi-Axion Models}

We now show that we can achieve inflation subject to the constraints of the WGC by generalizing to bi-axion (extra-)natural inflation, with 
two axions,  $A$, $B$ \cite{kimNillesPeloso, axion1, axion2, axion3, westphal, yangBai}. Consider the potential
\begin{equation}
	V = V_0 \left[ 1 - \cos \frac{A}{f_A} \right] + \tilde{V}_0 \left[ 1 - \cos \left( \frac{NA}{f_A} +\frac{B}{f_B} \right) \right], 
\label{eq:biax}
\end{equation}
where $N \in \mathbb Z$ by $A$-periodicity, following from its Nambu-Goldstone status.
  For sufficiently large $N \gg 1$, we get two hierarchical eigenmodes. At lower energies than the higher mass, the second 
   term enforces the constraint
\begin{equation}
	\frac{NA}{f_A} + \frac{B}{f_B} \approx 0. \label{integrateOut}
\end{equation}
Plugging back into $V$ gives an effective potential for the light mode, $\phi \approx B$,
\begin{equation}
	V_\text{eff}(\phi) = V_0 \left(1 - \cos \frac{\phi}{\feff} \right),  ~~\feff = N f_B.
\end{equation}
 
 This model is straightforwardly realized from 4+1 electrodynamics of two $U(1)$ gauge fields~\cite{yangBai},  $A_M$, $B_M$,~\footnote{Ref.~\cite{Cheung:2014vva} claims that there are additional constraints from the WGC in theories with multiple $U(1)$ fields, though this does not follow from our arguments. If there are $n$ $U(1)$'s all with a common coupling, then~\cite{Cheung:2014vva} claims that WGC bounds become stronger by a factor of $\sqrt{n}$, which is $O(1)$ in our examples.} with charges $(N,1)$ and $(1,0)$, and 4+1 masses less than $1/R$. Aharonov-Bohm effects analogous to \eqref{eq:aharanovBohm} then give rise to \eqref{eq:biax}, for effective 3+1 scalars, $A, B$ defined analogously to  \eqref{eq:WilsonLoop}, with $V_0 \sim \tilde{V}_0$ 
and $f_A = 1/(2\pi R g_A)$, $f_B = 1/(2\pi R g_B)$.
It is clear that the WGC,~\eqref{eq:WGC}, can be satisfied for both gauge interactions, with $f_A, f_B \ll  \mpl$, 
while still obtaining  $\feff \gg \mpl$, provided $N$ is large enough. 
Large $N$ also ensures that quantum tunneling of the fields through the potential barrier from the second term of Eq.~\eqref{eq:biax} is extremely suppressed.

But in non-renormalizable 4+1D QED, the UV scale of strong coupling (and EFT breakdown), $\Lambda_\text{gauge}$, falls rapidly as $N$ increases,
\begin{equation}
	\Lambda_\text{gauge} = \frac{8\pi}{N^2 g^2} \frac{1}{R}.
\label{eq:GaugeLambda}
\end{equation}
Minimally, both this cutoff and the WGC cutoff should be above the compactification scale, $1/R$, to remain in theoretical control. Given that for Natural Inflation, $\feff \gtrsim \sqrt{{\cal N}_\text{e-folds}} \mpl$,  it is easy to check that the bi-axion model can give parametrically large ${\cal N}_\text{e-folds}$ provided $N$ and $\mpl R$ are taken sufficiently large while keeping $N g$ fixed.

\section{Radius Stabilization} 

When 4+1 General Relativity is taken into account, $R$ is not an input parameter, but rather the expectation of a dynamical effective 3+1 (``radion") field, $\sigma(x^{\mu})$,
\begin{gather}
R = \mpl e^{\sqrt{\frac{2}{3}} \vev{\sigma(x)}/\mpl}.
\end{gather}
We show that $\mpl R \gg 1$ can arise naturally, and that the extra dimension is effectively rigid during inflation.
A suitable $\sigma$ potential can arise simply via Goldberger-Wise stabilization \cite{goldbergerWise}, in the case where the extra-dimensional circle is further ``orbifolded" in half, down to an interval. (This has the added benefit of projecting out the unnecessary 3+1 vector components of the gauge field, without otherwise affecting our earlier discussion.) The stabilization mechanism requires adding a 4+1
neutral scalar field, $\chi$.  The energy in this field depends on $R$, providing an effective potential for $\sigma$, 
\begin{gather}
	V_\text{radion} \sim m_\chi^2 M_5^3 \left( c_1 e^{2\pi R m_\chi} + c_2 e^{-2\pi R m_\chi} \right)  \nonumber \\
	\implies 2 \pi R \sim \frac{1}{m_\chi}, 
\end{gather}
where 
$c_{1,2} \sim {\cal O}(1)$ are determined by $\chi$ boundary conditions at the ends of the interval, and $M_5$ is the 4+1 Planck scale. 
Large $R$ clearly requires small $m_{\chi}$. This (and the small 4+1 cosmological constant that has been neglected above) can both be natural if the 4+1 ``bulk" spacetime preserves supersymmetry (to a high degree). The potential also gives the radion a mass, 
\begin{gather}
	m_\sigma^2 \sim \frac{1}{(2\pi R)^2} \gg H^2,
\end{gather}
so that it is not excited during and after inflation.

\section{Precision CMB Observables} 



CMB observables are sensitive to even small corrections to the inflationary potential. An attractive feature of the extra-dimensional realizations are that the structure of subleading corrections is controlled by the higher gauge symmetry. Eq.~\eqref{eq:aharanovBohm} shows that massive charges decouple exponentially from the potential, with the extra dimension effectively acting as a ``filter" of unknown UV physics, but they can have observable effects if not too heavy.  Since our effective theory has cutoffs on its validity given by the WGC, \eqref{eq:WGC},  and strong coupling in the UV, \eqref{eq:GaugeLambda}, in general new physics will appear by (the lower of) these cutoffs,  $\equiv \Lambda$. This may include new particles with 5D mass $M \approx \Lambda$ carrying charges $(n_A, n_B)$,  where each charge is plausibly in the range $|n| \lesssim N$. Such charges will create an Aharonov-Bohm correction to the potential, which after imposing the IR constraint, \eqref{integrateOut}, yields
\begin{gather}
	\delta V \sim V_0  \frac{(2\pi R M)^2}{3} e^{-2\pi R M} \cos \left(N n_B - n_A \right) \frac{\phi}{\feff}.
\label{eq:crap}
\end{gather}
If $N n_B - n_A \gg 1$, this ``higher harmonic'' gives a modulating correction to the slow-roll parameter  
$\epsilon \equiv \frac{\mpl^2}{2} \left( \frac{V'}{V} \right)^2$,
\begin{gather}
	\frac{\delta \epsilon}{\epsilon} = 2 (N n_B - n_A)  \frac{(2\pi R M)^2}{3} e^{-2\pi R M} \sin(N n_B - n_A) \frac{\phi}{\feff}.
\label{eq:deltaepsilon}
\end{gather}
For this to not obstruct inflation itself requires $\delta \epsilon/\epsilon < 1$. However, the parameter $\delta \epsilon/\epsilon$ also controls corrections to the temperature power-spectrum in the slow-roll limit, where the modulating part of the potential is almost constant during a Hubble time. Such periodic modulations of the inflationary potential have been searched for in the CMB data~\cite{Wang:2002hf, Pahud:2008ae,Flauger:2009ab,2011JCAP...01..026K,Easther:2013kla,Flauger:2014ana}, most recently motivated by the possibility of such signals in axion monodromy inflation~\cite{eva2,Flauger:2009ab,Flauger:2014ana}. These results place more stringent bounds, requiring  $\delta \epsilon/\epsilon \lesssim 1 - 5\%$, for $N n_B - n_A$ in a realistic range of $\sim {\cal O}(10-100)$.

Parametrically, it is easy to check that  $\delta \epsilon/\epsilon$ can be made arbitrarily small while still satisfying theoretical constraints, and consistent with large ${\cal N}_\text{e-folds}$. But this is accomplished at the expense of taking $\mpl R$ parametrically large. However, as seen in \eqref{eq:aharanovBohm}, $1/R$ sets the scale of $V_0$ in Natural Inflation, which is bounded by current observations. For example, we can fit the data, \eqref{eq:data},  with $N=42$, $g= 0.08$, $\mpl R = 8$. Then if we have new particles at the cutoff,  $M = \Lambda$ and charges $(n_{A}  \sim {\cal O}(N), n_B \sim {\cal O}(1))$, we have 
$\delta \epsilon / \epsilon \sim 3\% $. Of course, the charges at the cutoff may have a different pattern, and from \eqref{eq:deltaepsilon}, we are exponentially sensitive to order one uncertainties in determining $\Lambda$ from \eqref{eq:WGC} and~\eqref{eq:GaugeLambda}, but we see that our parametric success is also numerically plausible in the real world. Our estimates clearly motivate searching for such modulation in the CMB power spectrum.

\section{Tri-Axion Models}


Our discussion can be straightforwardly extended to tri-axion models~\cite{axion1, Higaki:2014pja, yangBai}, where smaller charge ratios are possible in the extranatural realization~\cite{yangBai}.  We find that such models can also satisfy the WGC, both parametrically and numerically in realistic models, with a higher and safer EFT cutoff. 
Consider 3 gauge fields $A, B, C$ and 3 particles with charges $(1,0,0)$, $(N_A,1,0)$, $(0,N_B,1)$. $N_A, N_B \gg 1$ implies only one light field, $\phi$, with
\begin{equation}
	\feff = \frac{N_A N_B}{2\pi R g_C}.
\end{equation}
We can now fit the data with smaller charges and lower corrections to the slow-roll parameter; e.g. taking $N_{A,B} =8$, $g_{A,B,C}= 0.12$, $\mpl R = 8$ we obtain $ \delta \epsilon /\epsilon \sim 3 \times 10^{-4}$.



\section{Chern-Simons Model}

The need for specific, large charges for light 4+1 matter may seem somewhat contrived. Arbitrary light charges would have effects similar in form to \eqref{eq:crap} but without Yukawa suppression, spoiling inflation. To explore this issue we modify our extra-dimensional approach so that these large quantum numbers become outputs of the model rather than fixed input parameters. For simplicity, we first focus on the single Maxwell field, 
$A_M$, and replace its coupling to
 explicit light charged matter by a Chern-Simons coupling to a non-abelian Yang-Mills (YM) gauge sector (say with $SU(2)$ gauge group), 
\begin{gather}
	\delta \L_\text{CS, 4+1} = \frac{N}{64\pi^2} \epsilon^{LMNPQ} G^a_{LM} G^a_{NP} A_Q.
\label{eq:5DCS}
\end{gather}
At this stage $N$ is still an input parameter,  its quantization enforced now by invariance under \emph{large} gauge transformations. 
In general, Chern-Simons couplings allow gauge fluxes to play the role of gauge currents; in this case YM fluxes act as an $A_M$ current. YM instantons can then replace the role of virtual Aharonov-Bohm effects. This is best seen by first passing to the 3+1 effective theory,
\begin{gather}
	\delta \L_\text{CS, 3+1} = \frac{N}{64\pi^2} \frac{A}{f} \epsilon^{\mu\nu\rho\sigma} G^a_{\mu\nu} G^a_{\rho\sigma}.
\label{eq:4DCS}
\end{gather}
This is very similar to the coupling of the Peccei-Quinn axion to QCD in order to solve the Strong CP Problem: upon YM  confinement~\cite{strongCP}
we obtain
\begin{gather}
	\delta \L_\text{$4D$ eff} = \hat{V}_0 \mathcal F\left( \frac{N A }{f} \right),
\end{gather}
where $\mathcal F$ is an order-one $2\pi$-periodic function replacing the second cosine in \eqref{eq:biax},  and $\hat{V}_0$ is set by the YM confinement scale. 
Similar generalizations $\mathcal F(NA) \to \mathcal F(NA + B)$ can replace \eqref{eq:biax}. In this way, we recover Natural Inflation via bi-axion or tri-axion models.

A virtue of the 4+1 Chern-Simons model is that it can be extended to 6+1 field theory with a Chern-Simons coupling, which may be written compactly in differential form notation as
\begin{gather}
	\delta \L_{CS, 7D} = \frac{1}{32 \pi^2} dA \wedge A \wedge G \wedge G,
\label{eq:7DCS}
\end{gather}
such that $N$ does not appear as an input coupling. Instead, we take the  6\textsuperscript{th}, 7\textsuperscript{th} dimensions to form a small 2-sphere, on which quantized 
$F = dA$ gauge flux can be  trapped. We will quantize about classical solutions with $N$ flux quanta,
\begin{gather}
	\oint_{S^2} F = \frac{N}{2\pi}.
\label{eq:flux}
\end{gather}
In this way, $N$ defines discrete selection sectors of the 6+1 theory,  a ``landscape'' of perturbatively stable vacuua. Plugging this condition into \eqref{eq:7DCS} reduces it to the 4+1 model, \eqref{eq:5DCS}. 

This basic mechanism can be extended to bi-/tri-axion models. For example, the second term of \eqref{eq:biax} can be produced if the $A$ field has a 6+1 Chern-Simons coupling as in \eqref{eq:7DCS} while the $B$ field has only a 4+1 coupling of the form in \eqref{eq:5DCS} to the same YM gauge sector. This could occur e.g.\@ if the $B$ field is localized to a 4-brane defect. In \cite{followUp} we will demonstrate that these 6+1 models are also parametrically controlled while being consistent with the WGC and ${\cal N}_\text{e-folds} \gg 1$. A key new feature in the analysis is the dynamical role $N$ plays in stabilizing the size of the 6-7 sphere. Note that obtaining many $e$-foldings of inflation does not require a very specific, ``tuned'' choice of $N$; large ${\cal N}_\text{e-folds}$ can in fact be generic within this landscape of solutions.

Let us  summarize. Black hole processes and properties provide a unique window into quantum gravity, placing tight constraints, such as the Weak Gravity Conjecture, on effective field theories of inflation. We have demonstrated that a parametrically large number of $e$-foldings of high-scale inflation can be realized by simple multi-axion generalizations of Extranatural Inflation, consistent with these constraints. The resulting models achieve large gravitational wave signals of $r \sim 0.1$ while remaining realistic and theoretically controlled, and predict potentially observable modulations of the scalar power spectrum. 



\begin{acknowledgments}
We thank Kaustubh Agashe, Nima Arkani-Hamed, Ted Jacobson, Marc Kamionkowski and Shmuel Nussinov for useful discussions and comments. The authors were supported by 
NSF grant PHY-1315155 and by the Maryland Center for Fundamental Physics. PS was also supported in part by NSF grant PHY-1214000. 
\end{acknowledgments}

\bibliography{InflationQGPRL}
\end{document}